\documentclass{epl}
\usepackage{nodisplaylabel,graphicx}

\newcommand{\twid}{\sim}

\newcommand{\gap}{\hspace{.4in}}

\newcommand{\gt}{\rightarrow}

\newcommand{\period}{\ \ .}
\newcommand{\comma}{\ ,\ }

\newcommand{\lsim}{\,\stackrel{<}{\scriptstyle \sim}\,}
\newcommand{\gsim}{\,\stackrel{>}{\scriptstyle \sim}\,}

\newcommand{\Angstrom}{\AA}

\newcommand{\ignore}[1]{}

{\begin{tabular}{#1}}%
{\end{tabular}}

{\end{list}}

{\end{list}}

{\ \\ XXXXX \hfill XXXXX #1 XXXXX\begin{equation}\label{#1}}%
{\end{equation}}

{\ \\ XXXXX \hfill XXXXX #1 XXXXX\begin{eqnarray}\label{#1}}%
{\end{eqnarray}}

{\begin{equation}\label{#1}}%
{\end{equation}}

{\begin{eqnarray}\label{#1}}%
{\end{eqnarray}}

\newcommand{\ie}{i.\,e.~}

{\begin{center} \section*{#1} \end{center}}

{\end{figure}} 

{\end{figure}}

{\end{figure}} 

\def\eqref#1{(\ref{#1})}

\newcommand{\aeff}{a_{\rm eff}}

\newcommand{\bstar}{b^*}

\newcommand{\fint}{f_{\rm int}}
\newcommand{\fext}{f_{\rm ext}}
\newcommand{\fel}{f_{\rm el}}

\newcommand{\HIC}{H_{\rm IC}}
\newcommand{\HPB}{H_{\rm PB}}

\newcommand{\HQB}{H_{\rm QB}}

\newcommand{\lB}{l_{\rm B}}
\newcommand{\lcharge}{l}
\newcommand{\lp}{l_{\rm p}}
\newcommand{\lsat}{l_{\rm sat}}
\newcommand{\lstar}{l^*}

\newcommand{\nc}{n_{\rm c}}
\newcommand{\ns}{n_{\rm s}}

\newcommand{\Neff}{N_{\rm eff}}

\newcommand{\Pstar}{P^*}

\newcommand{\qo}{q_0}
\newcommand{\qoeff}{q_0^{\rm eff}}

\newcommand{\rb}{r_{\rm b}}

\newcommand{\Ro}{R_0}

\newcommand{\lambdaGC}{\lambda_{\rm GC}}

\newcommand{\epsilonw}{\epsilon_{\rm w}}

\title{Srongly Charged Polymer Brushes}
\author{Ben O'Shaughnessy\inst{1} \and Qingbo Yang\inst{2}}
\institute{
  \inst{1}{Department of Chemical Engineering, Columbia University, New York, NY 10027, USA}
  \inst{2}{Department of Physics, Columbia University, New York, NY 10027, USA}
}


\ignore{
36.20.Ey Conformation (statistics and dynamics)
61.20.Qg Structure of associated liquids: electrolytes, molten salts, etc.
61.25.Hq Macromolecular and polymer solutions; polymer melts, swelling 
82.35.Rs Polyelectrolytes

61.41.+e Polymers, elastomers, and plastics
68.15.+e Liquid thin films
68.43.Mn Adsorption/desorption kinetics
82.70.Dd Colloids
} 

\begin{document}

\bibliographystyle{europhys}

\maketitle

\begin{abstract}

Charged polymer brushes are layers of surface-tethered chains.
Experimental systems are frequently strongly charged.  Here we
calculate phase diagrams for such brushes in terms of salt
concentration $\ns$, grafting density and polymer backbone charge
density.  Electrostatic stiffening and counterion condensation effects
arise which are absent from weakly charged brushes.  In various phases
chains are locally or globally fully stretched and brush height $H$
has unique scaling forms; at higher salt concentrations we find
$H\twid \ns^{-1/3}$, in good agreement with experiment.

\end{abstract}


\section{Introduction}

Charged polymer layers are a major concern of polymer science,
featuring in numerous applications such as colloid stabilization,
surface modification technologies, membrane preparation
\cite{pe_application_brush_PRL} and emerging biotechnologies
\cite{microarray_brush_PRL}.  A much-studied class of layer is the
polymer brush, an assembly of chains end-tethered to a substrate (see
fig.\ \ref{qb}).  Charged brushes are important in biosensor
technologies such as DNA microarrays \cite{microarray_brush_PRL} and
have provided model systems to study collective physics of charged
polymers at interfaces
\cite{tran:pss_brush,balastre:pebrush,tranauroy:cion,netzandelman:review,%
pincus:pebrush,borisov:pebrush_collapse,borisov:pebrush_diagram}.
Properties such as brush height have been measured by the surface
force apparatus, neutron scattering and other techniques
\cite{tran:pss_brush,balastre:pebrush,tranauroy:cion}.  The principal
theoretical frameworks have been self consistent field theory and
scaling theory \cite{netzandelman:review}.  Scaling theories
identified distinct brush ``phases''
\cite{pincus:pebrush,borisov:pebrush_collapse} and later a full
``phase'' diagram of brush types \cite{borisov:pebrush_diagram}.

Most of this theoretical work has addressed weakly charged brushes
where local Gaussian chain statistics are only weakly perturbed by
electrical forces.  Real polymers, however, are frequently strongly
charged.  The natural measure of this is the reduced backbone density
(``Manning parameter'') $\qo=\lB/l$ where $l$ is the (monovalent)
charge spacing and the Bjerrum length $\lB = e^2/\epsilonw kT = 7$
\Angstrom\ in water.  Strongly charged polymers include
single-stranded DNA ($\qo\approx 1.6$ \cite{tinland:ssdna_lp}) and the
widely studied polystyrensulfonate, PSS ({\it e.g.} $\qo \approx 2.8$
at 80\% sulfonation \cite{balastre:pebrush}).  When $\qo$ is not
small, two qualitatively new effects arise. (i) Strong repulsive
forces may stretch chains beyond the linear Gaussian regime into
rodlike configurations with end-to-end size of order the chain contour
length itself: the polymer size {\em saturates}. (ii) Manning
condensation \cite{manning:condensation}. When $\qo>1$, a rodlike
polymer attracts its own counterions so strongly that a fraction
condenses into a region close to the polymer, renormalizing the
effective Manning parameter to unity, $\qo\gt1$.  The implications of
(i) and (ii) have been explored theoretically in certain brush regimes
\cite{misravaranasi:satu_pebrush,naji:nonlinear_osb,hariharan:pebrush}.

The aim of this letter is to establish complete phase diagrams for
strongly charged brushes.  One of our motivations is the urgent need
to establish physics underlying DNA microarrays, a technology based on
DNA layers of enormous importance in health care, drug development and
basic biological research.  We will show that the saturation and
Manning condensation effects described above, (i) and (ii), lead to
brush phases whose scaling forms for brush height $H$ and counterion
layer thickness $D$ have no counterparts among weakly charged brushes
\cite{borisov:pebrush_diagram}.  Two brush classes are identified,
defined by the magnitude of the intrinsic persistence length,
$a$. Polymers in the ``strong'' class ($a>\lB$) are strongly perturbed
by charges, suffering length saturation for high enough backbone
charge density whereas for ``weak'' systems ($a<\lB$) Manning
condensation intervenes before saturation can onset.  Correspondingly,
each class has a distinct brush phase diagram.  The key point is that
typical experimental brush systems are ``strong'' and their chains
locally fully stretched, invalidating results based on flexible
Gaussian polymer concepts.  Brush height then depends on the
salt-dependent persistence length $\lp$ for a semi-flexible polymer.
The dependence of $\lp$ on salt concentration $\ns$ remains a highly
controversial question \cite{lp_brush_PRL,netzorland:lp}.  Here we
take $\lp \approx \xi$, a relation with considerable empirical support
\cite{tinland:ssdna_lp,nishida_brush_PRL}, where
$\xi=(8\pi\lB\ns)^{-1/2}$ is the Debye screening length.  For the
particularly experimentally important ``quasineutral'' brushes our
theory then predicts $H\twid \ns^{-1/3}$, in good agreement with
experiment \cite{tran:pss_brush,balastre:pebrush}. Interestingly, no
such agreement is obtained if one uses the often cited alternative
quadratic OSF form \cite{lp_brush_PRL}, $\lp\twid \xi^2$.  Thus brush
experimental data comes down rather firmly on the side of the
$\lp\twid \xi$ relation.

\section{Single Chains} 

To appreciate saturation and condensation effects, it is helpful to
first review results for a single charged chain of contour length $L$
and charge $Q = L/\lcharge$ \cite{schiesselpincus:cc_collapse}.
Throughout, we assume ideal Gaussian chain statistics in the absence
of charges.  For small charge, the chain size is $\Ro \approx
(La)^{1/2}$.  When $Q$ becomes large, internal repulsions exert a
stretching force $\fint \approx \lB Q^2 / H^2$
\cite{borisov:pebrush_diagram} (setting $kT=1$) where $H$ is the
extended chain size.  Balancing this with the elastic restoring force
$\fel \approx H/\Ro^2$ and using $\fint \rb \approx 1$ gives the
electrostatic ``blob'' size
                                                \begin{eq}{blob}
\rb \approx a (\lcharge/\lsat)^{2/3}		\comma\gap
\lsat \equiv (\lB a)^{1/2}			\period
                                                \end{eq}
For small scales $r<\rb$ statistics are essentially Gaussian, while on
larger scales the polymer is electrostatically stretched into a linear
string of blobs of length \cite{schiesselpincus:cc_collapse}
                                                \begin{eq}{singlechain}
H \approx \bstar (\lstar/\lcharge)^{2/3}	; \ \ \
	\bstar \equiv La/\lB  \comma \ \ \
			 \lstar \equiv \lB^2/a \period
                                                                \end{eq}
However, this result for $H$ cannot be correct for all charge
densities: from eq.\ \eqref{blob} when $\lcharge = \lsat$ the blob
shrinks to one persistence length $a$ and the chain size becomes $L$.
This is {\em size saturation}; further charging produces no further
extension.  (In this study, the window between onset of non-Gaussian
behavior and complete size saturation is ignored for simplicity.)

This suggests $H\approx L$ for $l>\lsat$. However, this reasoning
neglects Manning condensation.  Since the charge per blob is $\approx
\rb^2/al$, the string of blobs has an effective linear charge density
coarse-grained over the blob size equal to
                                                \begin{eq}{qo} 
\qoeff \approx (\lstar/\lcharge)^{1/3} \comma\gap 
			\lstar \equiv \lB^2/a \period
						     \end{eq}
Counterion condensation is triggered when $\qoeff=1$, {\it i.e.}, when
$l=\lstar$.  If we now consider a chain charging process (decreasing
$l$ from $l=\infty$), this leads us to define 2 classes of polymer:
(1) The weak class, $a<\lB$.  In this case $\lstar>\lsat$, so Manning
condensation onsets before saturation.  Thus eq.\ \eqref{singlechain}
for chain size remains valid until $l=\lstar$ when the polymer reaches
its maximum size, $\bstar$.  Charging beyond $\lstar$ generates just
enough condensate to pin the effective charge spacing to $\lstar$.
$H$ is pinned to $\bstar$ and saturation never occurs.  (2) The strong
class, $a>\lB$.  This includes most polymer species, {\it e.g.}\ PSS ($a
\approx 12$ \Angstrom) \cite{nishida_brush_PRL} and B-DNA ($a \approx
450$ \Angstrom) \cite{tinland:ssdna_lp}.  Saturation now occurs before
condensation can intervene, since $\lstar<\lsat$.  Thus eq.\
\eqref{singlechain} is valid until saturation at $l=\lsat$. For higher
charge densities, $H\approx L$.

We remark that the treatment of Manning condensation for weak systems
above is oversimplified.  One can go beyond this by treating the
string of blobs as a cylinder of radius $\rb$.  Adapting Ramanathan's
expression \cite{ramanathan:single_rod} for the potential at a charged
cylinder, $\psi(\qoeff) = 2 q \ln(\xi/\rb)$ with $q = \min(1,\qoeff)$,
a charging process then yields the free energy $F = (2Q - H/\lB)
\ln(\xi/\rb)$ for $\qoeff>1$.  Taking the derivative gives the
stretching force $\fint$ and balancing this with $\fel$ we find
$H=\bstar \ln(\xi/\rb)$, confirming that apart from the log factor
chain size is indeed pinned to $\bstar$.  Such logarithms, which arise
elsewhere ({\it e.g.} eq.\ \eqref{blob}) are ignored in the following
for simplicity.

\section{Brushes: Weak Systems}  

\begin{figure}
\includegraphics[width=7.5cm]{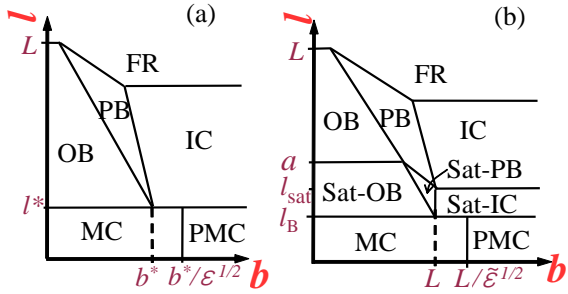}
\caption{\label{nosalt}
Brush phase diagrams for no added salt (grafting density $b^{-2}$,
backbone charge separation $l$, logarithmic axes). (a) Weak and (b)
strong systems.  OB: osmotic brush; PB: Pincus brush; IC: independent
chains; MC: Manning condensation; PMC: partial MC; FR: fully relaxed
chains.  Saturated phases are indicated by prefix
``Sat''. 
} \end{figure}

Consider now a brush of chains of the weak class, end-tethered to a
surface at density $b^{-2}$.  We begin with no added salt.  Our
results are shown as a phase diagram in fig.\ \ref{nosalt}(a).  The
upper portion, $l>\lstar$, is the weakly charged region previously
established by Pincus and Borisov et al.\
\cite{pincus:pebrush,borisov:pebrush_collapse,borisov:pebrush_diagram}.
We now review their results.  At low grafting densities lies the
Independent Chains (IC) phase where layer height $\HIC$ is essentially
the single chain result, eq.\ \eqref{singlechain}, and the height $D$
of the counterion layer ensuring overall neutrality is the
Gouy-Chapman length $\lambdaGC \equiv b^2/2\pi\lB Q$.  These
counterions exert a weak external stretching force $\fext \approx \lB
Q^2/b^2$ per polymer, less than the internal force $\fint$.  To the
left lies the Pincus brush (PB) where instead $\fext>\fint$. The IC/PB
boundary is $\fext \approx \fint$, or $\lcharge/\lstar \approx
(\bstar/b)^{3/2}$.  In the Osmotic brush (OB), the high grafting
density localizes counterions in the brush
\cite{borisov:pebrush_diagram}, unlike PB and IC where $D>H$. Thus the
PB/OB boundary is $\lambdaGC \approx \HPB$ or $\lcharge/\lstar \approx
(\bstar/b)^{4/3}$ \cite{borisov:pebrush_diagram}.  All 3 phases
coincide at the triple point $\Pstar = (\bstar,\lstar)$.  What happens
in the strongly charged region below $\lstar$?  The weakly charged
brush theory naturally raises a question it cannot answer.

The single chain analysis suggests the answer: the dependence on $l$
is removed by Manning condensation.  We find this is essentially
correct, using similar free energy calculations to those described for
single chains: for $l<\lstar$, brush height is fixed to $\bstar$ (to
within log factors) independent of $l$ or $b$.  This is the Manning
condensation (MC) phase in fig.\ \ref{nosalt} (a).  It has 2
sub-regions where $D \approx H$ (for $b<\bstar$) and
$D\approx\lambdaGC$ (for $b>\bstar$).  At the lowest grafting
densities the MC phase gives way to the partial Manning condensation
(PMC) region; here the counterions are so weakly attracted to the
surface ($D\gg H$) that condensate evaporation onsets
\cite{deshkovski:manning}.  We find onset at $b \approx
\bstar/\epsilon^{1/2}$, where the small parameter $\epsilon =
1/\ln(\bstar/\lB)$.  Deep into the PMC region, enough condensate is
lost that length saturation occurs, despite this being a weak system.
Finally, the boundary of the uppermost ``fully relaxed'' (FR) region
is defined by $H\approx\Ro$.

\section{Brushes: Strong Systems}  

Now we consider strong systems, $a>\lB$ (see fig.\ \ref{nosalt} (b)).
As expected from the single chain discussion, at high charge densities
new saturated phases appear (indicated by prefix ``Sat'') where brush
height reaches its maximum possible value, $H\approx L$.  The IC
region crosses over to Sat-IC at $\lcharge \approx \lsat$ (see eq.\
\eqref{blob}), while since chains in the OB have one charge per blob
\cite{borisov:pebrush_diagram} saturation is at $\lcharge\approx a$
(recall, blob size equals persistence length at saturation) giving the
Sat-OB phase.  Between these two lies the saturated Pincus brush
(Sat-PB).  Its boundary with Sat-OB is $\lambdaGC \approx L$ or $l/\lB
\approx (L/b)^2$; and with Sat-IC is $\fint \approx \fext$ or
$b\approx L$. Saturation effects shift the triple point $\Pstar$ to
$(L,\lB)$.

The essential feature of strong systems is that Manning condensation
onsets after chains have saturated and are rod-like, when $\qo=1$, \ie
$\lcharge=\lB$.  Analogously to the $\lcharge<\lstar$ phases for weak
systems, below $\lcharge=\lB$ lie the MC and PMC phases (fig.\
\ref{nosalt}, where $\tilde{\epsilon} \equiv 1/\ln(L/d)$ and $d$ is
polymer thickness.)

                                                  \begin{figure}
\includegraphics[width=8cm]{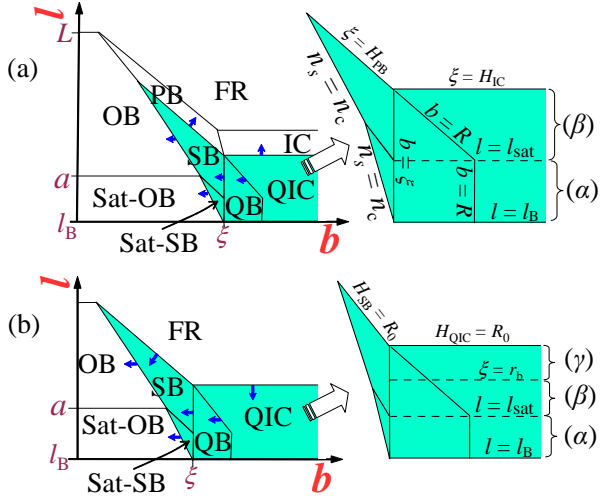}
\caption{\label{evolution} 
Brush phase diagram, strong systems with salt. The $\lcharge<\lB$
region (not shown) maps onto $\lcharge=\lB$.  Arrows indicate how
boundaries evolve with increasing salt concentration.  (a)
$L>\xi>\Ro$, (b) $\xi<\Ro$.  Salt-dominated regions (shaded) are blown
up at right.  QB (Quasineutral brush) and QIC (quasineutral
independent) are subdivided into bands $\alpha,\beta,\gamma$.  Since
$\rb<\Ro$, $\gamma$ does not exist for $\xi>\Ro$ (see text).
}
\end{figure}

\section{Salt Effects}  

For any real brush screening effects of salt must be considered.  For
brevity, we discuss strong systems only and the typical case $\xi<L$.
We find Manning condensation then renormalizes any $\lcharge<\lB$ to
$\lcharge\gt\lB$ with {\em no evaporation effects}.  Thus fig.\
\ref{evolution} shows only the portion of the phase diagram above
$\lcharge=\lB$; the region below is mapped vertically onto this line.
Figs.\ \ref{evolution}(a) and (b) show the two possible cases
$\xi>\Ro$ and $\xi<\Ro$.

Let us focus on the region affected by salt, shown shaded in fig.\
\ref{evolution} (compare to the no-salt case, fig.\ \ref{nosalt}).
This expands with increasing salt concentration as indicated and is
divided into two by the vertical line $b=\xi$.  To its left lie the
salt brush (SB) and saturated salt brush (Sat-SB) phases where
interactions are long-ranged and the stretching force is the
differential osmotic pressure \cite{borisov:pebrush_diagram} of ions.
The SB, established by weakly charged brush theories
\cite{pincus:pebrush,borisov:pebrush_collapse,borisov:pebrush_diagram},
saturates ($H\approx L$) at small $l$, crossing over to the Sat-SB
phase unique to strongly charged brushes.  Equating the salt
concentration $\ns$ to the brush counterion concentration $\nc \approx
Q/H b^2$ defines the salt/osmotic boundaries.  Thus the Sat-SB/Sat-OB
boundary is $l/\lB \approx (\xi/b)^2$.

                                                  \begin{figure}
\includegraphics[width=7.5cm]{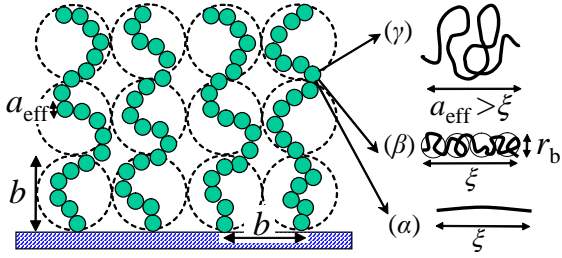}
\caption{\label{qb}
The quasineutral brush (QB) behaves as a neutral brush with effective
hard core monomer size $\aeff$ depending on polymer charge density and
salt concentration. There are 3 regimes, giving 3 bands in the phase
diagram, fig.\ \ref{evolution}.  Band $\alpha$: length saturation,
$\aeff\approx\xi$.  Band $\beta$: string of electrostatic blobs,
$\aeff\approx\xi$.  Band $\gamma$: overscreening ($\xi<\rb$), $\aeff \approx
\rb^3/\xi^2$.
}
\end{figure}

Moving to the right of $b=\xi$ one enters the quasineutral brush (QB)
phases. Since the screening length is less than the chain separation
$b$, interactions are now {\em short-ranged} so chains behave as self
and mutually avoiding polymers with a certain effective hard core
monomer size, $\aeff$, defined as the separation between any two chain
segments where their electrical interaction energy is order $kT$.
Thus we adapt the Alexander-de Gennes theory of neutral brushes
\cite{netzandelman:review} viewing each chain as comprising blobs of
size $b$ (see fig.\ \ref{qb}).  Depending on $l$, there are three
possible forms for $\aeff$: these are the three bands $\alpha, \beta,
\gamma$ in the phase diagram, fig.\ \ref{evolution}, with
corresponding structures shown in fig.\ \ref{qb}.  (1) In the
saturated $\alpha$ band, $\lcharge<\lsat$, polymers are locally
rod-like as discussed for single chains.  Flexibility is recovered on
scales above $\xi$ (recall, we take persistence length $\lp\approx
\xi$).  Thus $\aeff \approx \xi$ and there are $\Neff \approx L/\xi$
effective monomers per polymer. (2) The intermediate band,
$\beta$. Here the charge density is below saturation and locally the
polymer a string of electrostatic blobs of size $\rb$ (see eq.\
\eqref{blob}).  Thus $\aeff \approx \xi$ and $\Neff \approx
(\lsat/\lcharge)^{2/3} L/\xi$.  (3) In the overscreened $\gamma$ band,
screening penetrates individual blobs, $\xi<\rb$, reducing their
unscreened electrostatic energy which by definition equals $kT$.  Thus
$\aeff$ exceeds $\rb$.  From a Flory type calculation we find $\aeff
\approx \rb^3/\xi^2$ and $\Neff \approx La/\aeff^2$.

Borrowing the neutral brush result \cite{netzandelman:review}, the
brush height is $\HQB \approx \Neff\aeff^{5/3}b^{-2/3}$, giving
                                                \begin{eq}{hqb}
\HQB \approx \left\{
\begin{array}{ll}
(\xi/b)^{2/3}L \sim \ns^{-1/3} 
	\gap\gap(\alpha) \\
(\xi/b)^{2/3}L(\lsat/\lcharge)^{2/3} \sim \ns^{-1/3}
	\ (\beta\comma \ \gamma)
\end{array}
\right. .
                                                                \end{eq}
This is one of our principal conclusions.  The result for band $\beta$
has been derived previously by Borisov et al.\
\cite{borisov:pebrush_diagram}.  Note the identical scaling for
$\beta$ and $\gamma$ is coincidental.  

Finally, to the right of QB lies the quasineutral independent chains
(QIC) region whose boundary is $\HQB \approx b$.  The bands
$\alpha,\beta, \gamma$ remain, and chain size is the standard
self-avoiding result $R\approx \Neff^{3/5}\aeff$.  This agrees
qualitatively with the structure of a single charged chain found in
ref.\ \cite{netzorland:lp} though the predicted exponents differ.

We remark that if we assume the OSF form for persistence length
\cite{lp_brush_PRL}, $\lp \approx \lB (\xi/l)^2$, we find a
qualitatively similar phase diagram to fig.\ \ref{evolution} but now:
(i) between the Sat-SB and QB phases a very narrow saturated ``Nematic
Brush'' region appears where rodlike effective monomers nematically
order and (ii) new bands appear in the quasi-neutral phases. A crucial
difference is that the predicted salt dependence in the QB regime is
now $\HQB\twid \ns^{-1/2}$.

\section{Experiment}  

We conclude by comparing theory with two experimental studies of
charged brushes. (1) Tran et al.\ \cite{tran:pss_brush} used neutron
scattering to measure monomer density profiles of high density brushes
($13 \lsim b \lsim 29$ \Angstrom, $100 \lsim L \lsim 2000$ \Angstrom)
of PSS on silicon at sulfonation levels 35\% to 62\%. Using a
representative monomer size \cite{duboisboue:pss_lp} 2.1 \Angstrom,
this implies charge spacings $3.4 \lsim \lcharge$ $\lsim 6$ \Angstrom.
(2) Balastre et al.\ \cite{balastre:pebrush} studied less dense PSS
brushes on mica ($b\approx 100$ \Angstrom, $1000 \lsim L \lsim 1500$
\Angstrom, sulfonation 84\%-87\%, {\it i.e.}\ $\lcharge \approx 2.5$
\Angstrom) with the surface force apparatus.

Now PSS is a strong system ($a = 12$ \Angstrom) and both experimental
systems are strongly charged, $\lcharge < \lB$, thus belonging to the
line $\lcharge=\lB$ in the strong phase diagram, fig.\
\ref{evolution}.  Hence at low salt ($\xi>b$) we predict both systems
lie in the Sat-OB phase with fully stretched chains.  Indeed, the
observed low salt brush heights are $\gsim 0.7 L$ and $\approx 0.9 L$
in the Tran et al.\ and Balastre et al.\ studies, respectively.  This
is deep into the non-Gaussian regime
\cite{misravaranasi:satu_pebrush,naji:nonlinear_osb}, confirming the
chains are close to size saturation.

What happens when salt is added?  Our theory predicts the system in
effect moves along $l=\lB$ in fig.\ \ref{evolution}, passing from the
Sat-OB phase ($\xi\gg b$) to the $\alpha$ band of the QB phase
($\xi\ll b$) where chains are locally saturated.  The predicted brush
height $H$ is thus initially independent of salt concentration $\ns$,
then decaying as $\HQB \twid \ns^{-1/3}$ (eq.\ \eqref{hqb}).
Something very close to these predictions is seen in both experiments:
$H$ was roughly constant at low salt followed by a rather sharp
transition to an approximate power law $H \twid \ns^{-\zeta}$, with
$\zeta = 0.270$ observed in ref.\ \cite{tran:pss_brush} and $\zeta =
0.30$ to $0.33$ in ref.\ \cite{balastre:pebrush}.  The transition
occurred at $b/\xi \approx 3$ (ref.\ \cite{balastre:pebrush}) and
$b/\xi \approx 9$ (ref.\ \cite{tran:pss_brush}).  Since the Tran et
al.\ systems have very small $b$, the QB is realized only at very high
salt; we speculate the somewhat lower exponent $\zeta$ and higher
transition $b/\xi$ value for this data may originate from
non-electrostatic interactions (more important at high salt).

Throughout, we assumed Gaussian statistics, {\it i.e.}\ $\Theta$
solvents.  How would different solvent conditions affect the strongly
charged phases with saturation? (i) Good solvents are expected to
produce no qualitative change, since these phases are already locally
or globally almost fully expanded.  (ii) Very poor solvents and
ion-ion correlation effects have been predicted to induce brush
collapse \cite{borisov:pebrush_collapse,csajkaseidel:pb_md}.  For PSS,
for example, water is a poor solvent for the PS backbone.  However,
for high charge densities solubility is recovered
\cite{tran:pss_brush}; indeed, both PSS systems above exhibit highly
extended brushes at high charge density
\cite{tran:pss_brush,balastre:pebrush} as discussed.

In conclusion, we find strongly charged brushes exhibit unique phases
where chains are globally or locally almost fully extended into
rod-like configurations.  Systematic experiments
\cite{tran:pss_brush,balastre:pebrush} on strongly charged PSS brushes
find brush height $H \approx$ const.\ at low salt $\ns$, crossing over
to $H \twid \ns^{-\zeta}$ with exponents $\zeta \approx 0.3$.  It is
tempting to interpret this in a weakly charged brush picture
\cite{pincus:pebrush,borisov:pebrush_collapse,borisov:pebrush_diagram}
as a transition from the osmotic ($H\twid\ns^0$) to the salt brush
($H\twid\ns^{-1/3}$).  We find this is incorrect; the transition is in
fact from a saturated osmotic brush ($H\twid\ns^0$) to a quasineutral
brush with local saturation, for which we predict $H\twid\ns^{-1/3}$
provided the persistence length scales linearly with Debye length,
$\lp\twid \xi$ \cite{tinland:ssdna_lp,nishida_brush_PRL}.  The
observed exponent is clearly inconsistent with the OSF
\cite{lp_brush_PRL} form $\lp\twid\xi^2$ which would give a stronger
decay, $H\twid\ns^{-1/2}$.

\acknowledgments

This work was supported by the National Science Foundation, grant nos.\
DMR-9816374 and CHE-0091460.


\end{document}